 \documentclass[preprint2]{aastex}

\usepackage{epsfig}
\usepackage{epstopdf}

\usepackage{amssymb}

\shorttitle{The extinction curve}
\shortauthors{F. Zagury}

\begin{document}


\title{The 2200~\AA\ bump and the UV extinction curve}


\author{Fr\'ed\'eric Zagury}

\affil{Harvard University,  Cambridge, MA 02138, USA}
\affil{Fondation Louis de Broglie, 23 rue Marsoulan, 75012 Paris, France}
\email{fzagury@fas.harvard.edu}

\begin{abstract}
The 2200~\AA\ bump is a major figure of interstellar extinction.
Extinction curves with no bump however exist and are, with no exception, linear from the near-infrared down to  2500~\AA\ at least, often over all the visible-UV spectrum.
The duality linear versus bump-like extinction curves can be used to re-investigate the relationship between the bump and the continuum of interstellar extinction, and answer questions as why do we observe two different kinds of extinction (linear or with a bump) in interstellar clouds?
How are they related? 
How does the existence of two different extinction laws fits with the requirement that  extinction curves depend exclusively on  the reddening $E(B-V)$ and on a single additional parameter?
What is this free parameter?

It will be found that (1) interstellar dust models, which suppose the  existence of three  different types of particles, each contributing to the extinction in a specific wavelength range, fail to account for the observations;
(2) the  2200~\AA\ bump is very unlikely to be absorption by some yet unidentified molecule; 
(3) the true law of interstellar extinction must be linear from the visible to the far-UV, and  is the same for all directions including other galaxies (as the Magellanic Clouds).

In extinction curves with a bump the excess of starlight (or the lack of extinction) observed at wavelengths less than $\lambda=4000$~\AA\ is due to a large contribution of light scattered by hydrogen on the line of sight.
Although counter-intuitive this contribution is predicted by theory.
The free parameter of interstellar extinction is related to distances between the observer, the cloud on the line of sight, and the star behind it (the parameter is likely to be the ratio of the distances from the cloud to the star and to the observer).
The continuum of the extinction curve or the bump contain no information concerning the chemical composition of interstellar cloud.
\end{abstract}




\section{Introduction}
\begin{figure*}[]
\resizebox{2.2\columnwidth }{!}{\includegraphics{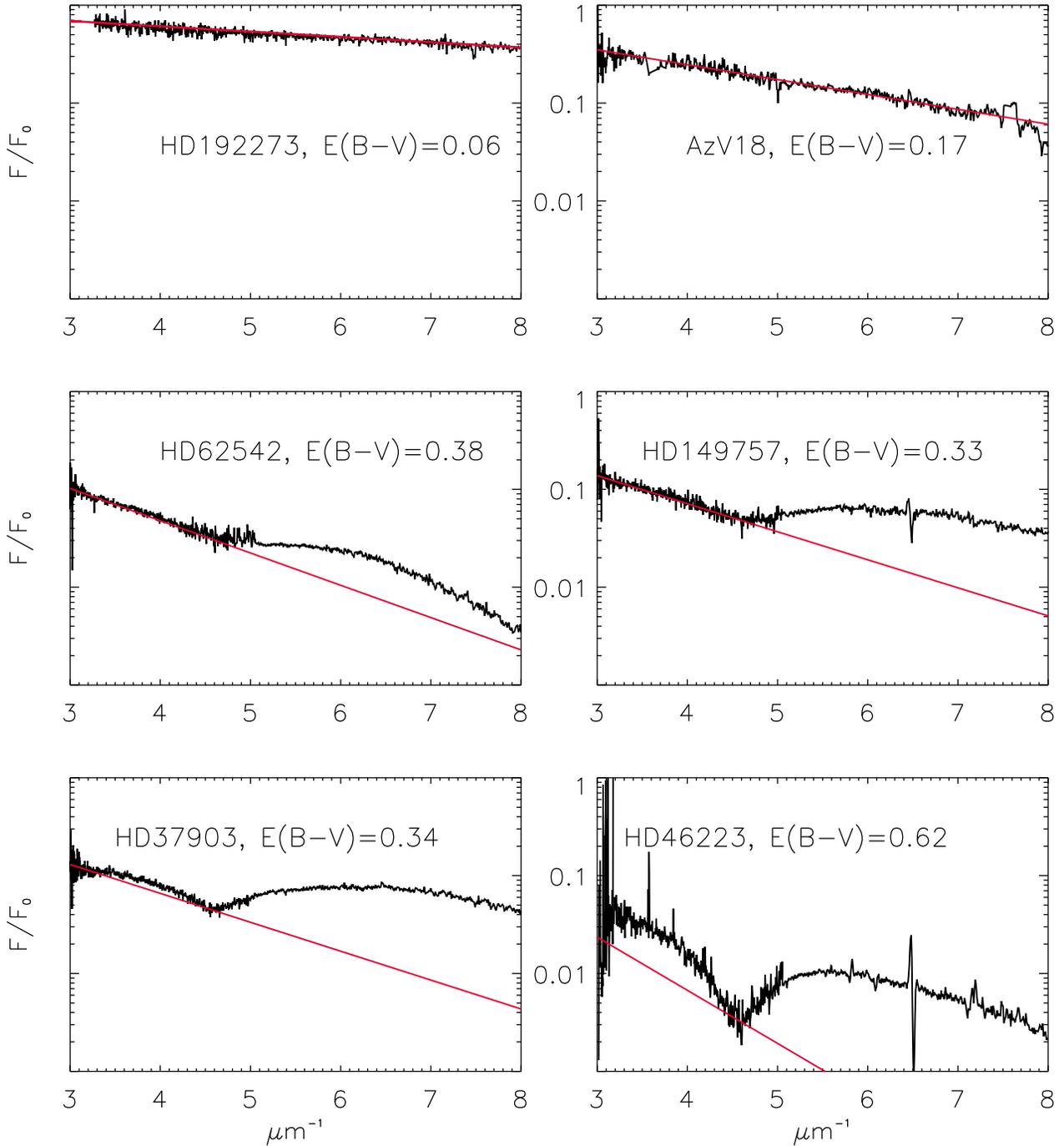}} 
\caption{The three types of observed interstellar extinction. The two upper $F/F_0$ curves (linear extinction laws, no bump) are found in low column density directions.
First occurrences of departure from linearity, middle spectra, arise in the far-UV, at slightly higher reddenings on the average.
Bottom spectra correspond to bump-like (CCM), most currently observed, extinction curves.
The observation of any reddened star, in the Galaxy or in the Magellanic Clouds, will fall into one of these three categories. There is no fixed threshold of reddening that separates one type of curve from the other.
The spectra have arbitrarily been normalized in such a way that the exponentials (red lines) be 1 at $1/\lambda=0$. 
Y-axis are logarithmic.
} 
\label{fig:fig1}
\end{figure*}
Early studies of interstellar extinction \citep{whitford58,nandy64} soon made it clear that the linearity of the visible extinction curve breaks towards the UV,  near 4300~\AA\ (the so-called Nandy knee \citep{wick98}).
The break is such that  extinction  in the near-UV is less than expected from the continuation of the visible extinction curve.
After 1970 observations from outside the atmosphere, especially those taken with the International Ultra-Violet Explorer (IUE) telescope, completed the curve and revealed an even more complex UV extinction.
Particularly spectacular is the absorption-like feature centered at 2175~\AA, the 2200~\AA\ bump, which interrupts the continuum of the extinction curve approximatively between 1800 and 2500~\AA.

The 2200~\AA\ bump, a central feature  of interstellar extinction,  has become a subject of research in its own right.
The wideness of the bump suggests a vibrational absorption band of a complex molecule.
No molecule has yet been formally  identified as the bump carrier but several have been proposed: PAH (polycyclic aromatic hydrocarbons) \citep{steglich10}, also thought to explain the far-UV rise of the extinction curve \citep{desert90}; 'GEMS' (glass with embedded metal and sulfides) \citep{bradley05}; 'buckyonions' \citep{li08}; nanodiamonds in glassy carbon shells \citep{yastrebov09}; polycrystalline graphite \citep{papoular09}; etc.
How these complex molecules manage to exist in sufficient abundance  in the interstellar medium is an unanswered and troubling question. 

This emphasis given to the bump has  eclipsed its relationship to the continuum of the extinction curve (Sect.~\ref{ccm}) and may limit its understanding. 
The variations of any two points on extinction curves with a 2200~\AA\ bump, whether they are in the bump region or not, are correlated \citep{ccm89,bondar06}.
This led \cite{ccm89} (hereafter CCM) to conclude that the shape of an extinction curve with a 2200~\AA\ bump depends on the reddening $E(B-V)$ and on one additional parameter only.
But the nature of this key parameter, essential to the comprehension of interstellar extinction and of the bump, and too quickly assumed to be the absolute V-band extinction to reddening ratio $R_V=A_V/E(B-V)$,  remains to be understood \citep{rv}.

The existence of directions with no bump (upper and middle plots of Fig.~\ref{fig:fig1}), for which the linear visible extinction\footnote{Here and farther in the text "linear" means that the extinction law is quasi-linear, that is it follows an $1/\lambda^p$ law, with $p$ close to 1 \citep{rv2}.} continues below 4000~\AA, down to the bump region or even in the far-UV, has also been neglected.
In the CCM framework no bump means no reddening.
The specific circumstances (low column density directions, interstellar matter close to the star, see Sect.~\ref{lin}), under which the bump tends to disappear represent another challenge: how can interstellar molecules which absorb at 2200~\AA\ be absent in some interstellar clouds and why, in this case, does the extinction law become linear over the whole spectrum?

The nature of the 2200~\AA\ bump will therefore not be fully understood unless its relationship to the continuum, and the absence of a bump in some observations, be clarified.
But regardless of the bump problem, the interpretation of the continuum of interstellar extinction curves is itself a source of  conflict.
First, the value of $R_V$ is independent of direction according to some studies, while other studies argue that it varies even on small angular scales \citep{rv}.
Second, as  already foreseen in CCM, the existence of different types of interstellar particles suggested by the different features of a UV extinction curve is  difficult to  reconcile with a one parameter  fit of normalized extinction curves.
In fact, all interstellar dust models rely upon a minimum of  six independent parameters (in addition to $E(B-V)$, see Sect.~\ref{cons}) instead of the single CCM one.

My purpose in writing this paper was to re-examine the large amount of data now available on interstellar extinction in light of the notable difficulties mentioned above.
In Sect.~\ref{obs}, I review the constraints observations impose on  the bump and the continuum of interstellar extinction curves.
Sect.~\ref{sca} to Sect.~\ref{bump} investigate the implications of these constraints.
\section{Observational constraints} \label{obs}
\subsection{The CCM fit and the degrees of freedom of interstellar extinction} \label{ccm}
CCM in 1989, and recently \cite{bondar06}, showed that the spatial variations of the extinction at two different wavelengths (in the infrared to UV wavelength range) are correlated:
plots of $A_\lambda-A_V$ versus $E(B-V)$ in different directions exhibit a linear relationship at all $\lambda$.
CCM  deduced that normalized extinction curves  constitute a set of curves of $1/\lambda$ which depend on only one, direction-dependent, parameter, $R_{ccm}$.
They derived an analytical fit for the curves and assumed that $R_V$ was its free parameter ($R_{ccm}=R_V$), which is certainly not the case (sect.~6 in \cite{rv}).
The question of the nature of the free parameter of interstellar extinction is still open.

The CCM fit of normalized extinction curves is a purely empirical relationship.
It is not always accurate (fig.~4 in CCM) and further applies to most but not all observations. 
Two reasons explain its inaccuracy: the relationship the fit establishes between the bump and $E(B-V)$, and its incapacity to reproduce linear extinction curves.
CCM  implicitly assumed  that  the shape of the 2200~\AA\ bump depends on $E(B-V)$ alone which can only be a first order approximation  \citep{savage75}.
The  CCM relationship only reproduces the set of extinction curves with a bump (no bump means $E(B-V)=0$), while extinction curves with no bump also exist, in the  Galaxy as well as in the Magellanic Clouds (Fig.~\ref{fig:fig1}). 
These curves, with no exception,  continue the linear visible extinction in the UV, at least down to the bump region.
Many are  linear over all the visible-UV spectrum and markedly distinct from the mean Seaton Galactic extinction law \citep{seaton79,lc}.
Linear cures over the whole spectrum can be included into a generalization of the CCM fit which still depends upon a single parameter (next Sect. and \cite{mc}).
\subsection{Linear extinction laws} \label{lin}
The absence or a reduction in the expected size of the 2200~\AA\ bump is  found in  two circumstances: low column density directions, and stars close to the obscuring material (shell stars, planetary nebulae).
Planetary nebulae have little or no bump, and when they do have a bump it is likely because of the presence of a foreground cloud on the line of sight (sects.~8-10 in \cite{z05}).

IUE observations of nebulae illuminated by nearby stars  lead to similar conclusions.
There is  no specific absorption at 2200~\AA\  in the UV observations of nebulae:
the spectrum of a nebula  deduces from the spectrum of its illuminating  star by the same linear law of $1/\lambda$ over the whole UV spectrum (including the bump region) and must continue the scattering law that should be found in the visible \citep{neb}.
When there is a  bump in the spectrum of a nebula illuminated by a nearby star, the bump exists in the same proportion in the spectrum of the star and  is due to foreground interstellar matter.

Very low extinction directions also have a linear extinction over the whole visible-UV spectrum (top plots of Fig.~\ref{fig:fig1} and  \cite{lc}).
Those directions exist in the Galaxy as well as in other galaxies; they are observed at much larger reddenings in the Magellanic Clouds \citep{mc}.

Extinction curves will nearly always be decomposed into a linear (over the whole spectrum)  and a CCM-like (with a bump)  components \citep{mc}.
The CCM part of the extinction curve and its free parameter are fixed by the size of the bump.
The bump is no more strictly related to the total reddening of the star, as requested by Savage's observations \citep{savage75}.
This decomposition provides a more general and considerably improved version of the CCM fit which still relies on a single parameter (in addition to $E(B-V)$).
\subsection{The three types of extinction} \label{3t}
Fig.~\ref{fig:fig1}  is a summary of  the three different types of UV extinctions which can be observed.
The $F/F_0$  representation\footnote{$F$ is the observed spectrum of the reddened star; $F_0$ the spectrum the star would have if it was not reddened. In practice, only $F_1$ ($\propto F_0$), the spectrum of a non or slightly reddened star of same spectral type, is available. 
The $F/F_0$ spectrum can be normalized from $F/F_0(1/\lambda \rightarrow 0)=1$ provided that $p$, the exponent of the quasi-linear law of the visible extinction, is known.
For the figures it has been arbitrarily assumed that  $p=1$.}, 
which is what observation most directly affords, was preferred to the traditional extinction curve [$A_\lambda $ or $E(\lambda-V)$, that is roughly $-2.5\log (F/F_0)$] for reasons to appear in Sects.~\ref{sca} to \ref{con}.
The y-coordinate is logarithmic, so exponentials are straight lines.

The two top spectra are low column density directions, one in the Galaxy, one in the Small Magellanic Cloud;
the extinction curve is linear over all the visible-UV wavelength range ($F/F_0$ is an exponential).
The bottom plots are traditional, most commonly observed curves with a bump.
Between  these two types, the two middle plots are observed in directions of intermediate reddening: the extinction remains linear down to the bump region, and diverges in the far-UV.
Extinctions of this type also exist in the Magellanic Clouds (Sk-69228 or Sk-70116, \cite{mc}).

Any extinction curve will inevitably fall into one of the three categories represented in Fig.~\ref{fig:fig1}.
There is no fixed threshold of reddening above which extinction turns over from one type to the other, from linearity to bump-like.
Linear extinction curves are more easily found among very low column density directions, curves with a bump towards large reddening ones, but the amount of reddening alone does not determine the type of extinction curve.
The star AzV18 in the SMC for instance,  with a reddening of 0.17~mag., has no 2200~\AA\ bump (Fig.~\ref{fig:fig1}).
A similar reddening in the Galaxy would most certainly result in a significant bump.
The transition from linear to bump-like extinction must be driven by a parameter other than $E(B-V)$, necessarily  the free parameter of interstellar extinction.
\subsection{Constraints on interstellar extinction theory} \label{cons}
Following the preceding sections, conservative requirements for a rational explanation of the observations on interstellar extinction would be that:
\begin{itemize}
\item interstellar extinction depends on the reddening along the line of sight and on one additional  parameter which varies from direction to direction;
\item extinction curves are linear (from the visible to the far-UV) for sufficiently low reddening;
\item extinction curves are linear when the interstellar matter is close to the star.
\end{itemize}
The expressions "\emph{sufficiently low}" in item 2 or "\emph{close to}" in item 3 are vague; they must be fixed by the free, yet unknown, parameter of interstellar extinction.

Existing dust models do not satisfy these conditions.
Any dust model will suppose the existence of at least three types of particles (large grains, bump carriers, small molecules for the far-UV rise), each responsible for the extinction in a particular wavelength range.
Since the weight of each of the near-UV, bump, and far-UV  features varies with direction, the relative proportion of each type of particles also needs to vary with  sight-line.
The size distribution of the large grains has to be truncated towards the small sizes, otherwise the UV extinction curve would continue the linear visible extinction law and there would be far more extinction in the UV than observed.
The threshold of the truncation cannot be the same whether the UV extinction curve is linear, partly linear, or bump-like.
It thus depends on direction.
It must also depend on reddening (second item above) and on distance from the stars (third item).

In practice the work of  \cite{fitz88} proves that any attempt to consider interstellar  extinction as the sum of three separate extinctions, which is what a dust model formally does, requires a minimum of five to six independent parameters in addition to $E(B-V)$, in contradiction with the first  item above.
\section{What alternative for the UV extinction curve?} \label{sca}
\begin{figure*}[t]
\resizebox{2.2\columnwidth }{!}{\includegraphics{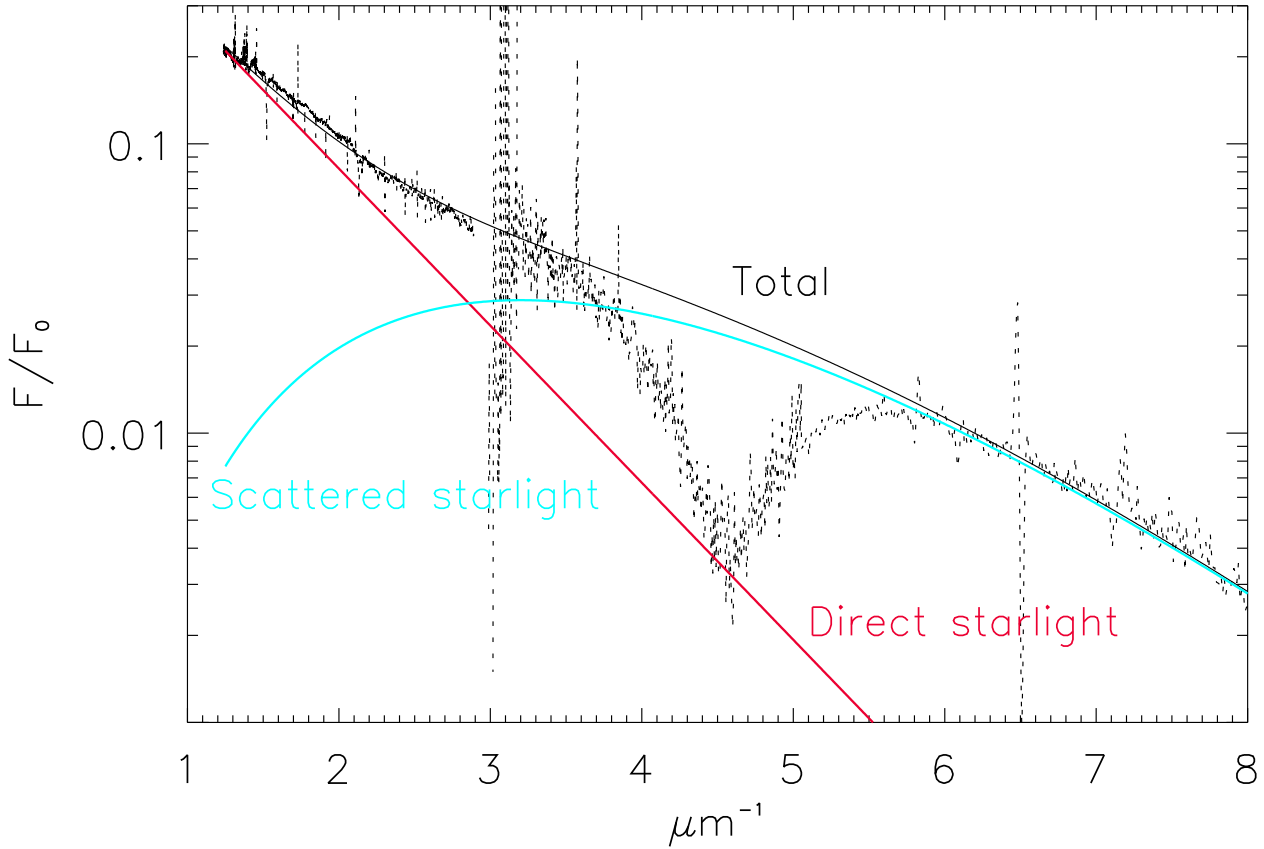}} 
\caption{Tentative decomposition of $F/F_0$ for HD46223 into its components of direct ($e^{-2E(B-V)}$, with $p=1$, $C_0=1$, Eq.~\ref{eq:sca}) and scattered starlight ($\propto e^{-2E(B-V)}/\lambda^4$, the bump region is omitted).
} 
\label{fig:fig2}
\end{figure*}
When extinction begins to depart from linearity in the UV, either in the near-UV (fig.~14 in \cite{nandy64}) or in the far-UV (middle plots of Fig.~\ref{fig:fig1}), the extinction curve $F/F_0$ systematically lays above the linear continuation of the visible extinction curve.
This deficiency in extinction has been attributed to a modification of interstellar grain properties, a change in the dust-size distribution, and the apparition of two new kinds of molecules.
The other and unique alternative is that the linear extinction of the direct light remains unaltered, and that what we observe is an additional contribution which has not  been accounted for yet.
This contribution must be provided by the cloud itself.

In the latter alternative the direct light from the star should follow the visible extinction law over the whole spectrum, and therefore scale  as $e^{-2E(B-V)/\lambda}$ ($\lambda$ in $\mu$m) \citep{rv}.
If light from the cloud, presumably scattered starlight, is added to the light from the star, the radiative transfer equation
\begin{equation}
\frac{F}{F_{0}}=e^{-\tau_\lambda}   
\label{eq:et}
\end{equation}
which presides at  extinction theory, is incomplete and does not apply anymore to the observation of reddened stars. 
Instead one must write
\begin{equation}
    \frac{F}{F_{0}}=C_0\left( 1+f_{s}\right)e^{-\alpha_pE(B-V)/\lambda^p}  \label{eq:sca}
\end{equation}
Parameter $\alpha_p$  depends on $p$ only and will be deduced from eq.~11 in \cite{rv2}.
$C_0$ should be 1 in absence of grey extinction, less than 1 otherwise.
$f_{s}$ represents the wavelength-dependent proportion of scattered starlight.
It must increase, as extinction does, with $E(B-V)$ and wave-number.
Both direct and scattered light components follow similar paths and need to be extinguished by the same amount of interstellar dust (the $C_0e^{-\alpha_p E(B-V)/\lambda^p}$ factor).

At very low reddening photons available for the scattering are scarce, scattered starlight is negligible and  direct light from the star should dominate over $f_{s}$ ($f_{s}\ll 1$). 
The extinction is then linear over the whole visible-UV wavelength range (upper plots of Fig.~\ref{fig:fig1}).
When extinction increases scattering will begin to show up in the far-UV first, as observed (middle plots of Fig.~\ref{fig:fig1}).
Increasing the extinction even more will lead to the merging of scattered starlight in the near-UV, and also, but to  lesser extent (Sect.~\ref{fsca}), in the visible.

The free parameter of interstellar extinction determines the departure from linearity of extinction curves.
It fixes the proportion of scattered light in the observed spectrum of a reddened star and must be included in the expression of $f_{s}$.
Since 2200~\AA\ bumps always appear with scattered starlight,  $f_{s}$ also contains all the information on the bump.

The hypothesis of a contribution of forward scattered starlight by the cloud to the spectrum of a reddened star  explains the observations of Fig.~\ref{fig:fig1} and satisfies  the two first items of Sect.~\ref{cons}.
\section{$f_{s}$'s continuum} \label{fsca}
$f_{s}$ increases with wave-number and reddening, contains the wavelength dependence of the bump, and depends upon a single parameter in addition to $E(B-V)$.
Fig.~\ref{fig:fig1} shows that $f_{s}$ can be larger than 1 in the UV (the intensity of the scattered starlight is then larger than the intensity of the star, even corrected for reddening).
I have initially suggested forward scattering by interstellar dust with a $1/\lambda$ dependence  \citep{neb}, but the recent paper by  \cite{hf} shows that a significant quantity of scattered light by gas is to be expected mixed to the light of a reddened star, should its distance from the interstellar cloud be large enough.

For a star whose spectrum  follows a linear extinction in the visible and has a large bump in the UV one should find in the visible, where $f_s\ll 1$,
\begin{equation}
    \left(\frac{F}{F_{0}}\right)_{vis} \approx C_0e^{-\alpha_pE(B-V)/\lambda^p}  \label{eq:fvis}
\end{equation}
In the far-UV, $f_s\gg 1$ and
\begin{equation}
    \left(\frac{F}{F_{0}}\right)_{uv} \approx  C_0f_{s}e^{-\alpha_pE(B-V)/\lambda^p}
     \label{eq:fuv0}
\end{equation}
$E(B-V)$ can be estimated from the visible part of $F/F_0$ and Eq.~\ref{eq:fvis}.
The wavelength dependence of $f_s$ in the far-UV will then be found from (Eq.~\ref{eq:fuv0})
\begin{equation}
     f_{s}\propto \left(\frac{F}{F_{0}}\right)_{uv}e^{\alpha_pE(B-V)/\lambda^p}    
     \label{eq:fuv}
\end{equation}
This method was applied to star HD46223 with 15~Mon (HD269698) as reference, two stars of same spectral type for which reliable visible and ultraviolet spectra were available \citep{hd}.
With  $p\approx1$ ($\alpha_p\approx 2$ with $\lambda$ in $\mu$m), it was found that outside the bump region the ratio of the two spectra is correctly represented by (Fig.~\ref{fig:fig2})
\begin{equation}
    \frac{F}{F_0}\approx C_0\left(1+\frac{c_s}{\lambda^4}\right) e^{-2E(B-V)/\lambda}  \label{eq:cs}
\end{equation}
In the far-UV $f_{s}$ can be approximated by a $1/\lambda^4$ power law.
The function $c_s$ depends weakly  on wavelength in the visible and in the far-UV, sharply in the bump region (Sect.~\ref{bump}).

This $1/\lambda^4$ dependence of $f_{s}$ can be checked in the far-UV from any data-set of stars with a bump.
Since from Eq.~\ref{eq:cs}
\begin{equation}
    \lambda^4 \left(\frac{F}{F_{0}}\right)_{far\,UV}\propto e^{-2E(B-V)/\lambda}  \label{eq:euv}
\end{equation}
the product of $\lambda^4$ times $F/F_0$ for any star with a bump should be close to an exponential of $-1/\lambda$, $e^{-2E_{uv}/\lambda}$, in the far-UV.
It is indeed observationally verified that $\lambda^4\times F/F_0$ follows an exponential with, as expected, $E_{uv}\approx E(B-V)$ \citep{fuv}.

The  $1/\lambda^4$ dependence of $f_{s}$ rules out  forward scattering by dust (scattering by dust would vary as $1/\lambda^p$, $p\approx 1$).
It implies Rayleigh scattering by particles small compared to the wavelength, presumably hydrogen which represents over 90\% of interstellar clouds.
Theory shows that a slab of gas in-between a star and an observer will, provided that distances be large, act as a lens and considerably enhance the apparent irradiance of the star \citep{hf}.
The coherence of the forward scattered starlight and the size of the Fresnel zones on astronomical distances are the two reasons which account for the importance of the light scattered by the slab.
The wavelength dependence ($1/\lambda^2$) of the irradiance of the scattered light indicated in \cite{hf}  does not match the $1/\lambda^4$ found above which I attribute to the too elementary treatment of the problem presented in the paper.
The role of the detection line (telescope + detector) and of distances was not investigated deeply enough and needs to be incorporated.
\section{The free parameter of interstellar extinction} \label{fpara}
 The \cite{hf} article shows that in order for scattering by gas to be significant distances need to be large.
Coherent scattered starlight vanishes when the gas is close to the star, as required by the third and last item of Sect.~\ref{cons}.

It is this dependence on distances that $c_s$ should include and which will determine the proportion of scattered starlight in the observed spectrum of a star.
The free parameter of the extinction curve must therefore be related to distances, presumably to the ratio of the distances from the star to the cloud and the cloud to the observer.
The function $c_s$ should also be proportional to the square of  H-atoms column density, that is to $E(B-V)^2$.

Extinction  in the direction of stars in the Magellanic Clouds can be separated into a linear and a CCM-like (with a bump) parts \citep{mc}.
The linear extinction is due to the reddening within the Magellanic Clouds, since dust in the Clouds is very close to the stars (compared to the sun-M.C. distance).
The CCM bump-like extinction on the other hand arises from Galactic cirrus on the line of sight: the scattered light component reaches a maximum when the star is infinitely far away \citep{hf}.
\section{The 2200~\AA\ bump} \label{bump}
Fig.~\ref{fig:fig1} shows that the bump is associated  to departure of extinction from linearity, thus to the presence of scattered starlight.
It was therefore noted at the end of Sect.~\ref{sca} that the analytical expression of the 2200~\AA\  bump had to be included in $f_{s}$ (or in $c_s$).
In Fig.~\ref{fig:fig2} the bump is indeed no deeper than the exponential decrease of the direct light from the star, as if  only the scattered starlight, and all of it, was extinguished.
Absorption by a molecule would not discern between the scattered and direct lights from the star; the bump is very unlikely to result from absorption, even by a complex molecule.
I see only two possible explanations for the feature, a minimum in the scattering cross-section of hydrogen or an interference/diffraction effect.
I found no ground for the former possibility.
The nature of the bump and its analytical expression should appear naturally in the exact derivation of the scattered light component. 

Noteworthy,  the first known diffuse interstellar band, the broad $\lambda$4430 band, is close to twice the central wavelength of the bump (the ratio is $\sim2.03$).
The DIB, if it was a second interruption of the scattered starlight, should be a few to over 10~\%  (fig.~2 in \cite{z05}) of the light of a reddened star\footnote{With a $1/\lambda^4$ dependence of the scattered light, the depths of the bump ($h_B$) and of the DIB ($h_D$) would be related by $h_D\sim 0.06 h_B/(1-h_B)$.}, which is the order of magnitude observation suggests.
\section{Discussion} \label{con}
A major obstacle to the understanding of interstellar extinction lays in the difficulty to establish good resolution extinction curves over the whole infrared to UV wavelength range.
Studies on interstellar extinction are for practical reasons compartmentalized between the infrared, the visible, and the UV.
Data for the three domains are of uneven quality  and have  little overlap.
In the UV the IUE database provides good resolution spectra for hundreds of stars, in all regions of the sky, but no such effort was made for the visible and the infrared.
Except  for \cite{divan54}'s spectral and \cite{nandy64}'s spectrophotometric studies in the visible, data in the visible/infrared consist in a few data-points that sample the spectrum.
Observations are also made from the ground and suffer from the difficulty to remove atmospheric extinction (which furthermore mimics interstellar extinction in the visible).
In the infrared, interstellar extinction is low, and uncertainties, or calibration problems, are probably often underestimated  \citep{rv2}.

The cross-section of interstellar grains shows a $1/\lambda^p$ ($p\approx 1$) dependence in the visible \citep{rv2}.
This extinction law probably extends to the infrared. 
In some directions linear extinction curves are also observed in the UV and must continue the linear  curve in the visible.
The determination of $p$ is an issue on which depends the value of $R_V$.
If, as observation indicates, linear extinction curves depend on  the reddening $E(B-V)$ alone, $p$ and $R_V$ must be the same constants  in all directions.
Their determination is up to date hampered by the absence of continuous extinction curves over a sufficiently large wavelength range.

The ultra-violet region, although the best documented, is also the most complex one.
Observed UV extinction curves  usually do not continue to be linear. 
New features appear, and $E(B-V)$ is no longer sufficient to fix the shape of an extinction curve.
Empirical works \citep{ccm89, bondar06,mc} have established that an additional parameter is enough to reproduce all extinction curves.
This parameter cannot be $R_V$ \citep{rv}.
Observation also shows that the transition of an extinction curve from its linear to non-linear regimes, whether it happens in the visible/near-UV (fig.~14 in \cite{nandy64}) or in the far-UV (middle plots of Fig~\ref{fig:fig1}), always appears as  less extinction than would be expected.

This paper attempted to prove that the only way to reconcile these observational constraints was to accept the idea that the light received from the direction of a reddened star, when its extinction curve has a 2200~\AA\ bump, is not direct light from the star alone, but contains a proportion of starlight scattered by the cloud on the line of sight.
An extinction curve then depends on the reddening and on the proportion of scattered starlight, that is on two parameters, as observation requires.
Since scattering, as extinction, increases with wave-number (in the present case it behaves as $1/\lambda^4$) it is less perceptible in the visible and even less in the infrared: extinction in these wavelength domains remains close to linearity and depends on $E(B-V)$ alone.
If the reddening  is low enough  scattered starlight is negligible: the extinction curve is linear from the infrared to the far-UV.
When extinction  increases scattered starlight will show up in the far-UV first, as observed.

Theory does predict \citep{vdh,hf} that a large proportion of scattered starlight should be mixed with direct starlight if the distances between the cloud, the star, and the observer, are large.
The free parameter of interstellar extinction is linked to these distances. 
The last requirement suggested by observation (Sect.~\ref{cons}), that the extinction becomes linear when interstellar matter is close to a star, is then also justified.
The bump observed in the extinction curves of stars in the Magellanic Clouds for instance, comes from cirrus in the Milky Way, not from interstellar matter in the Clouds.
The bump appears with the scattered light component only and does not extinguish more than the scattered starlight.
It is therefore very unlikely to be absorption by some kind of yet unknown molecule.

The free parameter of the extinction curve thus has nothing to do with interstellar grain properties.
The continuum of interstellar extinction does not contain any information on the composition of the particles which extinguish starlight.
It doesn't teach us anything about interstellar chemistry:
observations of interstellar extinction which diverge from linearity in the UV are a problem of optics, not of the chemistry of the interstellar medium.

Three main reasons have contributed to the non-recognition of the presence of scattered starlight in the spectrum of reddened stars.
A first reason is the standard representation of extinction curves (roughly $-2.5\rm{log}(F/F_0)\approx \tau_\lambda$, Eq.~\ref{eq:et}), adopted from the very beginning: scattered starlight is far more  easily identified on a $F/F_0$ plot (as in Fig.~\ref{fig:fig1}) than from the negative of its logarithm (as extinction curves are usually represented); furthermore, this is what observation directly provides.
A second reason is that the magnitude of the scattering  is also unexpected and counter-intuitive: it implies that more light can be received from the direction of a star when it is observed behind a  cloud of hydrogen than there  would be without the cloud.
The last reason is the lack of a reliable data-base of spectra in the visible, free from atmospheric extinction.
Plots of $F/F_0$ curves over the whole visible-UV spectrum could have  highlighted the relationship between the exponential decrease of interstellar extinction in the visible and its divergence with observation in the UV.

The consequences of the present work on interstellar extinction theory can be summarized as follows\footnote{Another consequence -the implication of a scattered starlight component in the spectrum of reddened stars for distance estimates- would need further investigations. Scattered starlight attenuates the impression of reddening. Photometric distances to reddened celestial objects may then be overestimated.}:
\begin{itemize}
\item The exact interstellar extinction law is linear over all the visible-UV spectrum. It is the same law for all directions, including directions in the Magellanic Clouds, and presumably for all galaxies.
\item Observations of the spectrum of reddened stars depend on two parameters: 1) the quantity of interstellar matter, measured by the reddening $E(B-V)$ (extinction depends on $E(B-V)$ and scattered starlight on $E(B-V)^2$);
2) the distances of  the cloud to the star and to the observer.
The ratio of these distances  fixes the importance of starlight scattered by the gas in the observed spectrum of a reddened star.
Interstellar extinction observations thus contain some (though probably loose \citep{hf}) information on distances.
\item $R_V$ is likely a constant. It is related to the best power law that describes linear extinction curves.
Spectral observations covering a large wavelength range will likely determine its value.
\item The bump is not an absorption feature. It corresponds to an interruption of the scattered light component in the spectrum of reddened stars. It may be related to some of the diffuse interstellar bands.
\item Neither the continuum of the extinction curve nor the 2200~\AA\ bump can be used to constrain the chemistry or the molecular composition of the interstellar medium.
\end{itemize}
The exact, physically meaningful, fit of interstellar extinction curves remains to be established.
\section*{Acknowledgments}
This work was supported by an Arthur Sachs  fellowship. I am grateful for the hospitality and resources provided by Harvard  University.
\bibliographystyle{model3-num-names}
{}
\newpage

\end{document}